# Revealing Grain Boundaries and Defect Formation in Nanocrystal Superlattices by Nanodiffraction

Nastasia Mukharamova[1], Dmitry Lapkin[1], Ivan A. Zaluzhnyy[1,2,+], Alexander André[3], Sergey Lazarev[1,4], Young Y. Kim[1], Michael Sprung[1], Ruslan P. Kurta[5], Frank Schreiber[6,7], Ivan A. Vartanyants[1,2,*], and Marcus Scheele[3,7,*]

[1]*Deutsches Elektronen-Synchrotron DESY, Notkestrasse 85, D-22607 Hamburg, Germany*

[2]*National Research Nuclear University MEPhI (Moscow Engineering Physics Institute), Kashirskoe shosse 31, 115409 Moscow, Russia*

[3]*Institute of Physical and Theoretical Chemistry, University of Tübingen, Auf der Morgenstelle 18, 72076 Tübingen, Germany*

[4]*National Research Tomsk Polytechnic University (TPU), pr. Lenina 30, 634050 Tomsk, Russia*

[5]*European XFEL GmbH, Holzkoppel 4, D-22869 Schenefeld, Germany*

[6]*Institute of Applied Physics, University of Tübingen, Auf der Morgenstelle 10, 72076 Tübingen, Germany*

[7]*Center for Light-Matter Interaction, Sensors & Analytics LISA+, University of Tübingen, Auf der Morgenstelle 15, 72076 Tübingen, Germany*

[+]*Present address: Department of Physics, University of California San Diego, La Jolla, CA 92093, USA.*

## Abstract

X-ray nanodiffraction is applied to study the formation and correlation of domain boundaries in mesocrystalline superlattices of PbS nanocrystals with face-centered cubic structure. Each domain of the superlattice can be described with one of two mesocrystalline polymorphs with different orientational order. Close to a grain boundary, the lattice constant decreases and the superlattice undergoes an out-of-plane rotation, while the orientation of the nanocrystals with respect to the superlattice remains unchanged. These findings are explained with the release of stress on the expense of specific nanocrystal-substrate interactions. The fact that correlations between adjacent nanocrystals are found to survive the structural changes at most grain boundaries implies that the key to nanocrystal superlattices with macroscopic domain sizes are strengthened interactions with the substrate.

Superlattices of inorganic nanocrystals (NCs) are often viewed as large-scale analogues to crystalline lattices of atoms.[1–5] In line with this analogy, homogenous NC ensembles crystallize in common closed-packed structures, such as face-centered cubic (*fcc*), body-centered cubic (*bcc*), body-centered tetragonal (*bct*) or hexagonal close-packed (*hcp*) arrangements.[6–12] Such ordered superlattices are held together by directional forces between adjacent NCs, which span a wide spectrum from weak van-der-Waals interactions to ionic and covalent bonds.[6,13–16] Anisotropic, facet-specific interactions invoke orientational order of NCs into mesocrystalline assemblies with a global angular correlation between the superlattice and the atomic lattices of its NCs.[17–20] When these interactions are weakened, e.g. due to stress induced by polydisperse NCs or by uniaxial strain applied to the superlattice, defects occur which lead to the manifestation of grain boundaries and polycrystallinity.[21–23] Since the surface of NCs consists of different facets with different polarities, ligand binding motifs and strengths, defects in the superlattice form predominantly along the direction with the weakest binding strength.

One of the most common defects in NC superlattices are twin boundaries, which have previously been studied by electron microscopy, electron diffraction or small-angle X-ray scattering.[8,21,24–29] Furthermore, NC superlattices are prone to other defects such as point, line, planar or volume defects.[21,24,30–32] A detailed understanding of their origin is expected to improve the design of NC superlattices with tailored mechanical, electric and optical properties.[33,34] Equally important is their role as a model for grain boundary formation in general and the nature of the spatial transition between these. Specifically, in view of the previously found correlations between the orientation of individual NCs and the superlattice, are these correlations preserved through grain boundaries, and, if so, in which way?[12,13,20] How does the structure change close to the grain boundaries? Addressing these fundamental questions on structure formation would also shed light on the intriguing question to which degree superlattices of NCs serve as a model for atomic crystalline systems and where this analogy ends.[2]

Here, we use angular X-ray cross-correlation analysis (XCCA) in conjunction with a nanofocused X-ray beam to reveal the structure and orientational order in superlattices of PbS NCs linked by oleic acid molecules near grain boundaries.[20,35–37] We find that the superlattice forms an *fcc* structure and that the lattice constant is homogeneous within a single-crystalline domain. Close to the edges, it decreases by 5-10 %, which is often accompanied by a rotation of the superlattice. We determine two different angular correlations between the superlattice and the atomic lattices of the NCs. This highlights the greater flexibility of the interparticle attractions in *fcc* superlattices of NCs compared to *bcc* assemblies which exhibit mostly a single angular correlation.[12,20,27] Our results enable a deepened understanding of the origin of defects in NC superlattices, highlight the role of orientational order in this respect and serve to tailor the mechanical properties of NC-based materials.

## Results and discussion

### Definition of crystalline domains and their orientation

The X-ray scattering experiments were performed at the Coherence Beamline P10 of the PETRA III synchrotron facility (see Methods for details). The PbS mesocrystal sample was scanned with 250 nm resolution, and small angle as well as wide angle X-ray scattering (SAXS and WAXS) were simultaneously measured. **Figure 1a,b** displays an X-ray scanning image of a polycrystalline superlattice of PbS NCs, utilizing the positions of superlattice (SL) peaks at $q^{SL} = 0.083$ Å$^{-1}$ (**Figure 1a**) and of the atomic lattice (AL) peaks at $q^{AL} = 2.12$ Å$^{-1}$ (**Figure 1b**) (see Supplementary Materials for more details). Throughout this manuscript, all reflections, planes or crystallographic directions referring to superlattice of the PbS NCs will be denoted with the index "$_{SL}$" and from the atomic lattice will be denoted with the index "$_{AL}$". Spatial positions of the sample with the same angular orientation of peaks in SAXS (**Figure 1d-h**) and WAXS (**Figure 1i-m**) diffraction patterns are indicated by the same color and specified by individual diffraction patterns. Grey color stands for areas with scattering signal but without a well-defined angular orientation (see Supplementary Materials for details). A comparison with an optical micrograph of the sample (**Figure 1c)** shows very good resemblance between the spatially distributed scattering intensity and the real-space image.

From the color code, the domain structure of the superlattice of NCs with single-crystalline grains and areas of 50-100 µm$^2$ is easily visible. We determine five typical patterns for both SAXS (**Figure 1d-h**) and WAXS (**Figure 1i-m**) which are sufficient to categorize all SAXS and WAXS patterns.

The integrated SAXS patterns (**Figure 1d-h**) contain many orders of SL peaks and can be indexed according to an *fcc* structure with the cell parameter $a^{SL} = 150$ Å (see also Supplementary Materials, **Figures S3, S4 and Table S1** for further details). The nearest-neighbor distance in this case is 106 Å. This value is consistent with the NC size of 68±5 Å measured by optical spectroscopy and the length of oleic acid of 19 Å (on each NC) as detailed in the Supplementary Materials (**Figures S1, S2**). The narrow size of all diffraction spots indicates that these parameters are uniform over each domain, and we note that we find excellent agreement with these q-values in all domains of the sample (see Supplementary Materials). For the SL, we observe either the $[001]_{SL}$ or the $[110]_{SL}$ direction along the surface normal as the two dominant orientations. In the rare case of the orange-colored SAXS domain, the $[301]_{SL}$ direction is perpendicular to the surface.

At wide angles, we observe X-ray scattering from the {111} and {200} planes of the AL with rock-salt structure at $q_{111}^{AL} = 1.84$ Å$^{-1}$ and $q_{200}^{AL} = 2.12$ Å$^{-1}$, respectively, which allows us to unambiguously determine the orientation of the AL in each WAXS domain. Specifically, in the yellow, green, and orange WAXS domains (**Figure 1i,j,k**), we detect 111$_{AL}$ and 200$_{AL}$ peaks, while in the purple WAXS domain (**Figure 1m)** we find two 200$_{AL}$ peaks, and a single 200$_{AL}$ peak in the red WAXS domain (**Figure 1l**) (see also **Figure S3** in Supplementary Materials). These patterns correspond to two dominant orientations of the AL: the yellow, green and orange

WAXS domains exhibit the $[110]_{AL}$ direction perpendicular to the sample surface, while in the red and purple WAXS domains the $[100]_{AL}$ direction is perpendicular to the sample surface. Three domains, namely the red, orange and violet, exhibit nearly identical spatial dimensions in SAXS and WAXS. In contrast, the yellow WAXS domain coincides with two SAXS domains – green and yellow. From here on, we will focus on a discussion of the SAXS domains.

**Determination of the NC orientation in the SL**

To determine correlations between the superlattice structure and the orientation of its NCs, we apply an XCCA approach (see Supplementary Materials for details).[36,37] We calculate the two-point cross-correlation functions (CCF) for all five domains according to

$$C(q_1, q_2, \Delta) = \langle I(q_1, \varphi) I(q_2, \varphi + \Delta) \rangle_\varphi, \qquad (1)$$

where $I(q,\varphi)$ is the intensity at $(q,\varphi)$ point of the diffraction pattern and $\varphi$ is the angular coordinate around a diffraction ring. We correlate the intensities of the momentum transfer $q_1 = q_{200}^{SL} = 0.83$ Å$^{-1}$, $q_2 = q_{200}^{AL} = 2.12$ Å$^{-1}$ for the yellow and red domains, and $q_1 = q_{311}^{SL} = 0.143$ Å$^{-1}$, $q_2 = q_{200}^{AL} = 2.12$ Å$^{-1}$ for the green, purple and orange domains.

In **Figure 2**, we determine the relative orientations of the NCs in the sample. The figure is structured as follows: the first row (**Figure 2a-e**) shows the CCFs. The second row (**Figure 2f-j**) contains the corresponding WAXS and SAXS patterns integrated over all diffraction patterns of each of the five domains. The third row (**Figure 2k-o**) displays the simulated real space orientations for each domain of the SL based on the geometrical interpretation of the scattering data. The simulated CCFs (blue dashed curves) in **Figure 2a-e** are based on these real space structures (see Supplementary Materials for details). The good agreement between experimental and simulated CCFs supports the structural interpretation used to index **Figure 1d-h** and serves to understand all further scattering patterns in this work, including individual patterns at grain boundaries. Except for the red and purple domains, we observe four correlation peaks for each domain. For the red domain, there are only two peaks in the CCF due to the low intensity of one pair of SAXS peaks (see **Figure 2a**). This may be explained by a significant (5-10°) tilt of the SL with respect to the sample surface. However, the relative positions of the SAXS and WAXS peaks in this domain (**Figure 2f**) are similar to the yellow domain (see **Figure 2h**). For the purple domain, there are eight peaks in the CCF (**Figure 2b**) as this domain is characterized by the presence of two $<200>_{AL}$ and four $<311>_{SL}$ reflections (see **Figure 2g**). The CCFs for all domains except for the orange one are symmetric with respect to $\Delta = 0°$, indicating the symmetry of the angular position of the NCs with respect to the $<110>_{SL}$ directions. We have marked the axis of symmetry for the red domain at 0.3° under the assumption that its CCF resembles that of the yellow domain. The axis of symmetry for the orange domain is at 42°, which means that the NCs are not positioned symmetrically in the $(30\bar{1})_{SL}$ plane. This is further illustrated in **Figure 2o** by the asymmetric real space structure in this orientation.

**Compression of the superlattice near domain boundaries**

Now we turn our attention to the study of the SL structure close to domain boundaries. **Figure 3a** shows variations in the value of the $q_{200}^{SL}$ momentum transfer within the entire sample. We find that the average value of $q_{200}^{SL}$ in the center of each domain far from its boundary is 0.083 Å$^{-1}$ (see **Figure S3f-j**), while near the edge its value increases up to 0.10 Å$^{-1}$. This indicates a compression of the superlattice as one approaches the domain boundary. To illustrate this, we select two scans through the boundaries of the purple (Scan S1) and orange (Scan S2) domains. At each point of the scans, we calculate the positions of the $<200>_{SL}$ reflections shown in **Figure 3b,c** and find an increase of the momentum transfer by 10% and 5%, respectively. We observe the same trend for the peaks of the <311> family (see **Figure S7** in the Supplementary Materials). Evidence for a contraction of the superlattice near domain boundaries is also found in a previous report based on electron microscopy (see Figure 1a therein).[38]

**Rotation of the superlattice near domain boundaries**

In **Figure 4**, we exemplarily analyze the changes in the orientation of the superlattice and of the NCs near grain boundaries observed along scans S2 and S3 marked in **Figure 1a,b**. Scan S2 represents the approach from a mesocrystalline domain towards an area without any scattering (orange into white), while scan S3 is an example for a grain boundary between two mesocrystalline domains (green into yellow). In scan S2, the superlattice undergoes an out-of-plane rotation around the $[010]_{SL}$ axis by 22°, and is simultaneously tilted by 7° (**Figure 4a-d**) as schematically illustrated in **Figure 4e,f**. The rotation angles were obtained by simulation of each individual diffractions pattern from **Figure 4a-d** (see **Figure S10** in the Supplementary Materials). At the same time, the atomic lattices of all NCs exhibit an out-of-plane rotation around the $[\bar{1}00]_{AL}$ direction as evidenced by the emerging $11\bar{1}_{AL}$ Bragg peak (**Figure 4b,c,** see also **Figure S11**). The changes in the relative intensities of the $020_{AL}$ and $11\bar{1}_{AL}$ peaks indicate the rotation of the AL around the same $[010]_{SL}$ axis, thus we conclude that the angular correlation is most likely preserved in this example. In the previous section we already described the shrinking of the lattice along this scan (see **Figure 3c**). Therefore, remarkably, although the orientation and lattice spacing of the superlattice changes drastically close to the edge of the sample, the angular correlation of the superlattice with the NCs is preserved until the scattering signal vanishes. In scan S3 (**Figure 4g-j**), the WAXS pattern remains practically unchanged, while the SAXS pattern exhibits a 90° rotation over a distance of 4 µm. This sequence of diffraction patterns can be rationalized as an out-of-plane rotation of both, the SL and AL, around the $[\bar{1}\bar{1}0]_{SL}$ and $[001]_{AL}$ axes (which are collinear in the green and yellow domain) by 90° as detailed in **Figure 4k,l**. Importantly, the angular correlation between the SL and AL is thus preserved across this grain boundary.

While the SAXS and WAXS patterns of scans S1 and S4-S11 are provided in the Supplementary Materials (**Figure S8 and S9**), we note the following general trends: Near a grain boundary, the superlattice always experiences an out-of-plane rotation with typical values between 8-14°. For the orientation of the NCs, a clear trend is less obvious since the relatively broad Bragg reflections with a full width at half maximum of 15° (see **Figure S6**, **Table S2**) make small out-of-plane rotations difficult to monitor. However, even despite this obstacle we frequently observe the disappearance of Bragg peaks of the AL while the SL rotates. This indicates either an out-of-plane rotation of the NCs or a fainting scattering signal due to a thin material coverage at the edges of the sample. None of the scans show a pronounced in-plane rotation of the NCs. Thus, the typical grain boundary is characterized by an out-of-plane rotation of the mesocrystalline unit cell by 8-14°, a compression of the lattice constant by 5-10% and a preservation of the angular correlation between the superlattice and its constituting NCs. Domains of different angular correlation (e.g. yellow *vs.* red or green *vs.* purple) are separated by extended areas without long-range order of the superlattice and/or no material between these domains.

## Discussion

From the XCCA analysis in **Figure 2f-j**, we obtain two different angular correlations, which apply to all domains of the sample (**Figure 5**). In the first configuration ("Conf1", **Figure 5a**), we observe the collinearities $[110]_{SL}\|[100]_{AL}$ and $[001]_{SL}\|[001]_{AL}$. The second configuration ("Conf2", **Figure 5b**) is characterized by the collinearities $[110]_{SL}\|[100]_{AL}$ and $[001]_{SL}\|[011]_{AL}$. We note that a single rotation of all NCs by 45° around the $[100]_{AL}$ transposes Conf1 into Conf2. While the purple and red domain are characterized by Conf1, the yellow, green and orange domain are examples for Conf2. To verify this for the orange domain, we simulate the SAXS and WAXS patterns for the superlattice in **Figure S2c,h,m** and find that taking the incident beam directions along the $[30\bar{1}]_{SL}$ axis reproduces the SAXS pattern of the orange domain as well as the relative scattering intensities of the $200_{AL}$ and $111_{AL}$ Bragg peaks. Indicated by the relatively small intensity of the $111_{AL}$ Bragg peak, the $[30\bar{1}]_{SL}$ and $[\bar{1}0\bar{1}]_{AL}$ axes are not exactly collinear, but the difference is negligible when taking into account the calculated misorientation of the NCs (see Supplementary Materials). The geometrical relationship between the red and purple domain as well as the yellow, green and orange domain is indicated with the corresponding colors in **Figure 5a** and **Figure 5b**, respectively.

Using an X-ray beam with 100 µm footprint, Li *et al.* have previously reported diffraction patterns from *fcc*-PbS NC superlattices exhibiting both configurations Conf1 and Conf2 simultaneously.[10] This was attributed to a mixed unit cell of the superlattice consisting of two groups of NCs with the two different angular correlations. In contrast, we exclusively find domains with uniform angular correlation. A possible explanation for these apparently contradicting findings is the much smaller X-ray foot print utilized by us, enabling the analysis of single-crystalline domains of the superlattice. We suggest that the stability of the two configurations is based on the maximization of ligand-ligand interactions between adjacent $\{100\}_{AL}$ and $\{111\}_{AL}$ facets separated by the nearest neighbor distance. In an SL with *fcc* structure, each NC is coordinated twelve-fold via its nearest neighbors along the twelve $<110>_{SL}$ directions. We detail this in **Figure 5c,d** for both configurations by displaying a $(1\bar{1}1)_{SL}$ plane for each, which contain the central NC and six nearest neighbors. If a nearest neighbor directly faces the central NC with a $\{100\}_{AL}$ or $\{111\}_{AL}$ facet, this is indicated with red and blue ligand molecules, respectively. Otherwise, the ligands between adjacent NCs are omitted for clarity. Conf1 stands out in that all twelve $<110>_{SL}$ directions exhibit such ligand interactions, namely four $\{100\}_{AL}$-$\{100\}_{AL}$ and eight $\{111\}_{AL}$-$\{111\}_{AL}$ interactions (see **Figure 5c and Figure S12a-c**). In contrast, Conf2 exhibit only ten such interactions, respectively, all of which occur exclusively between $\{100\}_{AL}$ facets (see **Figure 5d and Figure S12d-f**). Since these facets are roughly $10^6$ times more reactive than $\{111\}_{AL}$ facets, we believe that the smaller number of total interactions compared to Conf1 is compensated by the larger number of specific $\{100\}_{AL}$-$\{100\}_{AL}$ interactions.[39–41]

Recent molecular dynamics simulations have revealed a rich phase diagram for superlattices of NCs depending on the ligand coverage, the particle shape and size, the ligand length as well as the amount of residual solvent trapped in the superlattice.[42] For a partially solvated superlattice, for instance due to trapped, residual solvent molecules, a truncated octahedral particle shape and a diameter of 6.8 nm, the computed phase diagram predicts the presence of an *fcc* structure with three distinct angular correlations ("O3-*fcc*"). We emphasize that the structures observed by us in the present work should be classified as two different types of "UA-*fcc*", i.e., an *fcc* lattice with a uniform angular correlation of all NCs. In this regard, we note that another phase described by Fan *et al.* ("UA-*bct*") is virtually identical to Conf1, apart from a very small tetragonal distortion. Small distortions may be difficult to observe experimentally due to the effect of inhomogeneous shape and size of the NCs. Furthermore, it is noteworthy that for PbS superlattices with *fcc* structure, polymorphs with only two distinct angular correlations in one domain have been found experimentally.[10,43] Similar results were obtained for Si NC superlattices with *fcc* structure.[44] These findings were explained with facet-specific interactions between the substrate and the NCs, directing the NCs into a preferred orientation towards the substrate regardless of the orientation of the superlattice. However, from the five different AL orientations in **Figure 1b**, only two are characterized by $\{100\}_{AL}$-substrate interactions (red and purple) and in no case $\{111\}_{AL}$ interactions with the substrate are observed. The remaining orientations exhibit interactions between $\{110\}_{AL}$ facets and the substrate, but the area of these facets is typically negligible such that the expected interaction energies should be small. It is noteworthy that $\{100\}_{AL}$-substrate interactions are observed only for Conf1 and $\{110\}_{AL}$-substrate interactions only for Conf2. To conclude, the effect of facet-substrate interactions cannot be excluded (in particular for Conf1), but a dominant effect of ligand-ligand interactions is more consistent with our observations in explaining the polymorphism of the *fcc* structure.

We suggest that facet-substrate interactions play a far more important role in the formation of grain boundaries. Each grain boundary studied by us in **Figure 3** and **Figure 4** is accompanied by an out-of-plane rotation of the superlattice and a preserved angular correlation with the atomic lattices of the NCs. Thus, the type and number of ligand-ligand interactions does not change in the vicinity of this structural defect, in contrast to the interactions with the substrate. Based on the observation that all grain boundaries are further characterized by a change of the lattice constant (**Figure 3**), we believe that the predominant driving force for the out-of-plane rotation of the superlattice is the reduction of stress or strain. A strategy towards mesocrystalline PbS NC superlattices with larger coherent domains should therefore aim at reducing stress/strain in the superlattice - *e.g.* by a narrower size distribution – as well as maximizing interactions of the NCs with the surface of the substrate to prevent out-of-plane rotations. We anticipate that this may be achieved by coating the substrate with a self-assembled monolayer of a molecular species that interacts strongly with the ligands of the NCs.

## Conclusion

We have identified two distinct mesocrystalline polymorphs of PbS nanocrystal superlattices with face-centered cubic structure, both of which exhibit a rigid iso-orientation of the nanocrystals with the superlattice. We explain this polymorphism with the number of facet-specific ligand-ligand interactions, which are extraordinarily large for both structures. Boundaries between single-crystalline domains occur upon an out-of-plane rotation of the superlattice and the nanocrystals under full preservation of their angular correlation. This out-of-plane rotation is probably caused due to stress or strain in the superlattice as evidenced by a 5-10 % compression of the lattice constant during the rotation. We suggest that the key to nanocrystal superlattices with improved long-range order are stronger particle-substrate interactions as the particle-particle interactions are advantageously strong.

## Methods

The X-ray diffraction experiment was performed at the Coherence Beamline P10 of the PETRA III synchrotron source at DESY. The nanodiffraction end station GINIX was used to focus an X-ray beam with energy E = 13.8 keV ($\lambda$ = 0.898 Å) down to 400 x 400 nm$^2$ size with KB-mirrors. The depth of the X-ray focus was about 0.5 mm. The sample was positioned perpendicular to the incoming X-ray beam as shown in **Figure 6**. An area of 30 x 30 μm$^2$ was scanned to analyze the spatial variations of the structure of the samples. Within this scanning region, 14641 diffraction patterns were collected on a 121 x 121 raster grid with about 250 nm step size in both directions perpendicular to the incident beam. Each diffraction pattern was collected with an exposure time of 0.3 s to stay below the threshold of radiation damage, which was assessed by repeating the scanning procedure several times on the same position of the sample. A two-dimensional detector Eiger 4M (2070 x 2167 pixels with 75 x 75 μm$^2$ size) was positioned downstream at a distance of 41 cm from the sample and shifted so to have the transmitted beam close to a corner. With this geometry a part of reciprocal space in wide angle scattering was accessible and we were able to detect the scattering signal from the NC SL as well as from PbS AL simultaneously.

PbS nanocrystals of about 6.8 nm in size capped with oleic acid were synthesized following Weidman *et al.*[45] The NCs were drop-cast from a hexane:octane (9:1) solution onto 5 x 5 mm$^2$ Si frames with a 500 x 500 μm$^2$ window consisting of a 50 nm-thick Si$_3$N$_4$ membrane (PLANO).

## Acknowledgements

We thank T. Salditt for providing nano-focusing instrument (GINIX) support at the P10 beamline. This work was supported by the DFG under grants SCHE1905/3, SCHE1905/4 and SCHR700/25. This work was supported by the Helmholtz Association's Initiative and Networking Fund and the Russian Science Foundation (Project No. 18-41-06001).

## Supporting Information

(S1) Comparison of the lattice parameter with the nanocrystal diameter; (S2) Transmission electron micrograph of the nanocrystals used in this study; (S3) Determination of the structure of the superlattice; (S4) Simulation of the SAXS patterns of all five domains; (S5) Details of the X-ray cross-correlation analysis; (S6) Determination of the nanocrystal disorder; (S7) Additional evidence for a compression of the superlattice near grain boundaries; (S8) and (S9) Diffraction patterns of all further scans described in the main part; (S10) and (S11) Simulation of SAXS and WAXS diffraction patterns of scan 2; (S12) All facet-specific ligand-ligand interactions for Conf1 and Conf2

## Materials & Correspondence

*To whom correspondence and material requests should be addressed:

ivan.vartaniants@desy.de

marcus.scheele@uni-tuebingen.de

## Contributions

I.A.V., R.K., F.S., I.A.Z., S.L. and M.S. designed the study; N.M., D.L., I.A.Z., S.L., Y.Y.K. and F.S. conducted the study; N.M. and D.L. analyzed the data; A.A. prepared the samples; M.Sprung provided the experimental infrastructure; N.M., D.L., I.A.V., F.S. and M.S. wrote the paper.

## Competing interests

The authors declare no competing interests.

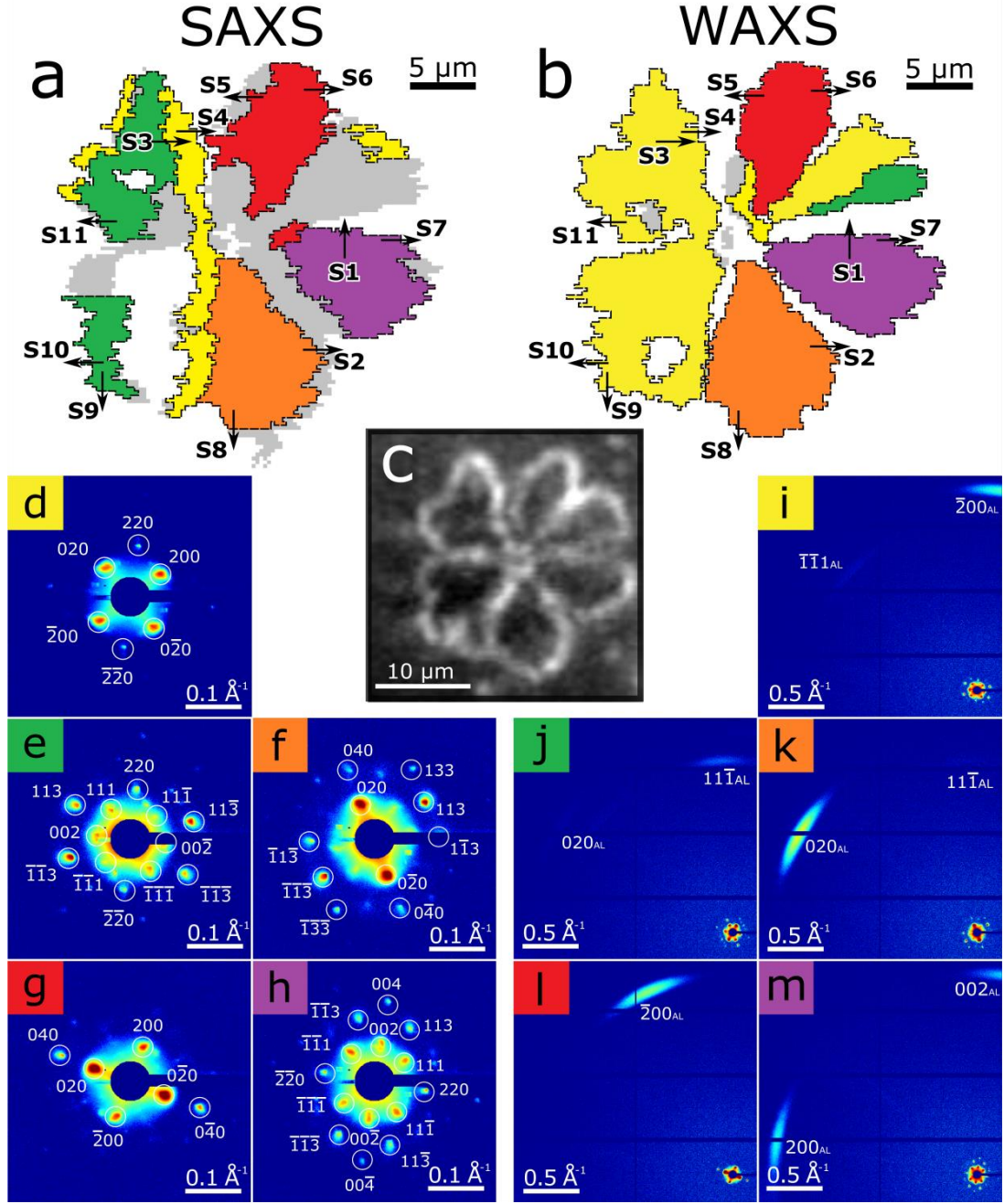

**Figure 1.** Spatially resolved maps of a mesocrystalline superlattice of PbS NCs, showing the domain structure. Each color corresponds to different orientations of the peaks at $q^{SL} = 0.083$ Å$^{-1}$ for the SL (**a**) and the peaks at $q^{AL} = 2.12$ Å$^{-1}$ for the AL (**b**). Grey color stands for areas with SAXS or WAXS scattering present but without well-defined orientation, while white areas correspond to parts without any scattering. Black arrows refer to specific scans across the grain boundaries. (**c**) An optical micrograph of the same sample. (**d-m**) Average SAXS (**d-h**) and WAXS (**i-m**) diffraction patterns with the corresponding peak indexing for each colored domain are shown. The peaks are indexed under the assumption of an *fcc* structure for the SL and a rock-salt structure of the NCs.

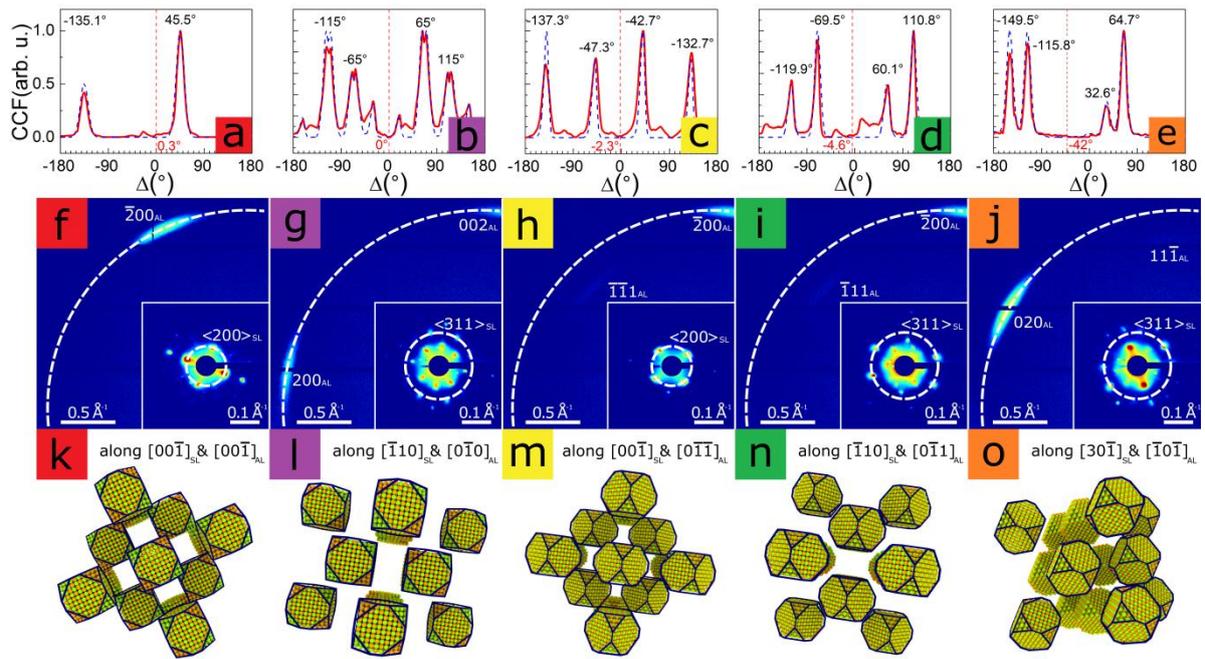

**Figure 2. (a-e)**: Calculated CCFs (red lines) based on the experimental data and simulated CCFs (blue dashed lines) based on the model structures shown in (k-o) for all five domains, using the *q*-values described in the text. The analyzed *q*-values are also indicated by the white dashed lines in the SAXS and WAXS patterns (**f-j**). (**f-j**): Averaged WAXS intensity of diffraction patterns corresponding to each domain. Enlarged SAXS patterns are shown in the lower right corners (Note: The center of the WAXS pattern does not coincide with the center of the SAXS pattern due to the different scale). (**k-o**): Real space models of the superlattice and its constituting NCs based on the SAXS and WAXS patterns and CCFs for all five domains.

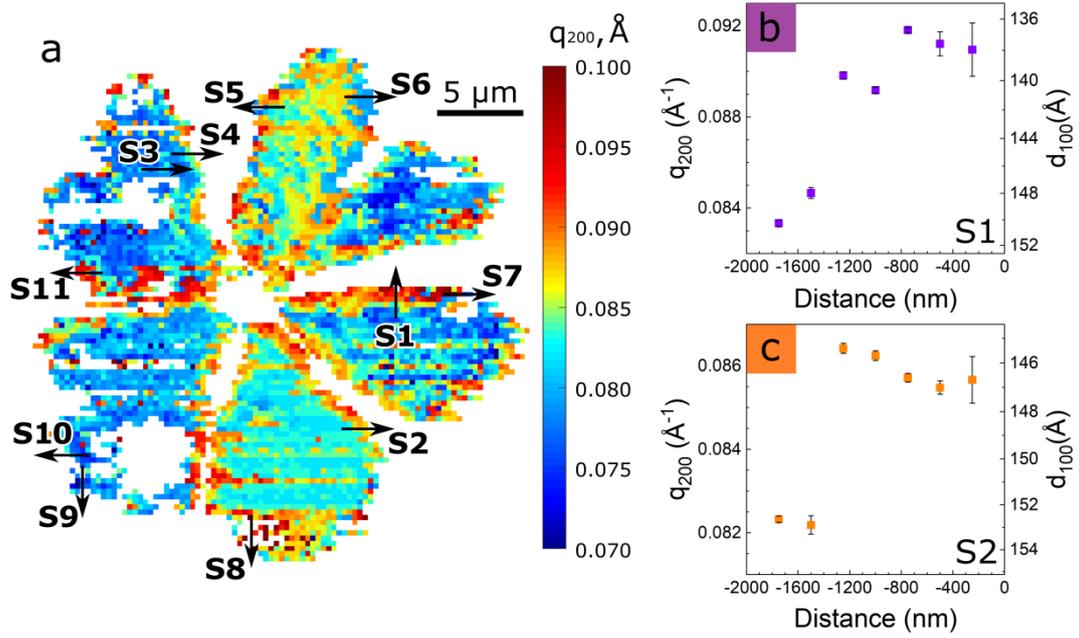

**Figure 3. (a)** Spatial map of the momentum transfer value for $<200>_{SL}$ reflections. The color code quantifies the value of $q_{200}^{SL}$. Black arrows refer to specific scans discussed in the text. **(b-c)** Spatial variation in $q_{200}^{SL}$ between the bulk and the edge of domains for scans S1 (**b**) and S2 (**c**).

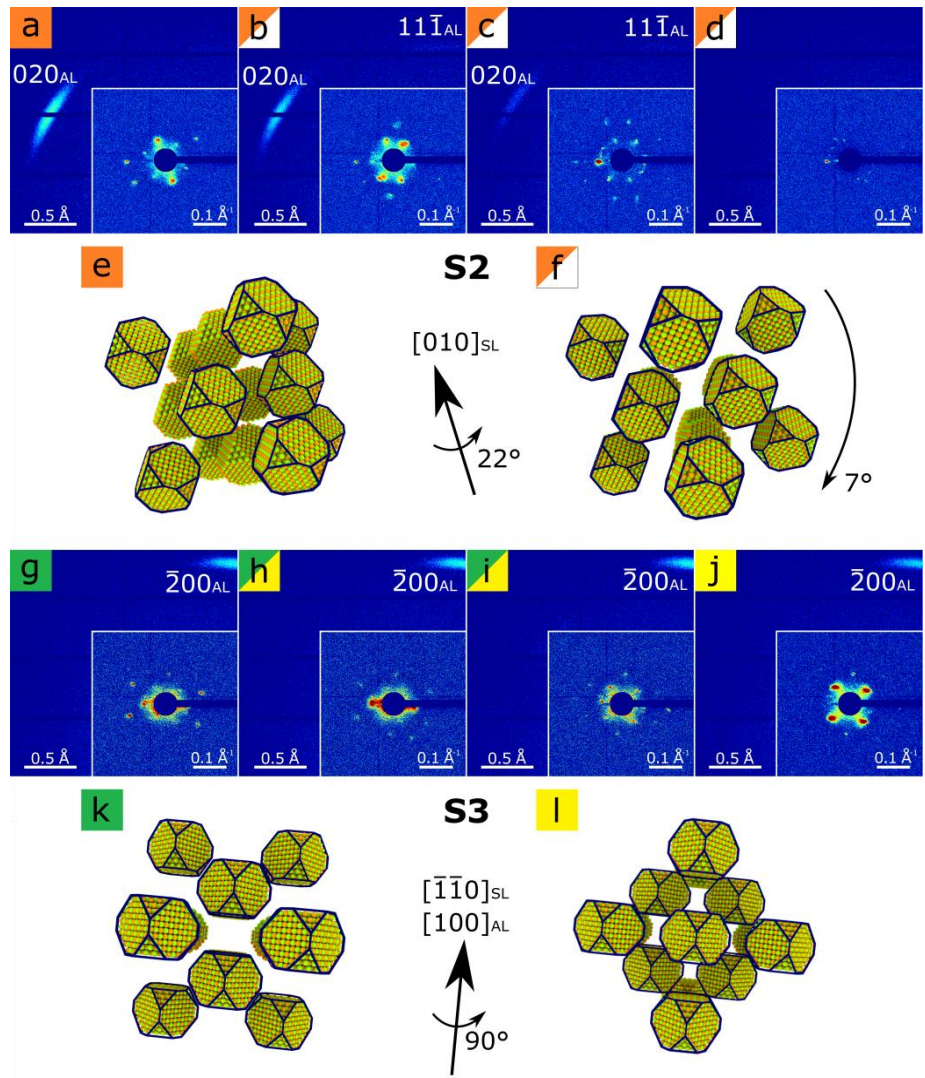

**Figure 4.** (**a-d**) Scan through the border of the purple domain (scan S2 in **Figure 1a-b**) with 500 nm step size. The unit cell rotates by 22° around the [010]$_{SL}$ direction and is additionally tilted by 7° out-of-plane (**e, f**). (**g-j**) Scan through the border between green and yellow domains of the SL (scan S3 in **Figure 1a-b**) with 1000 nm step size. The unit cell rotates by 90° around the [$\bar{1}\bar{1}0$]$_{SL}$ and [100]$_{AL}$ directions as indicated in the schematic real space representations (**k, l**).

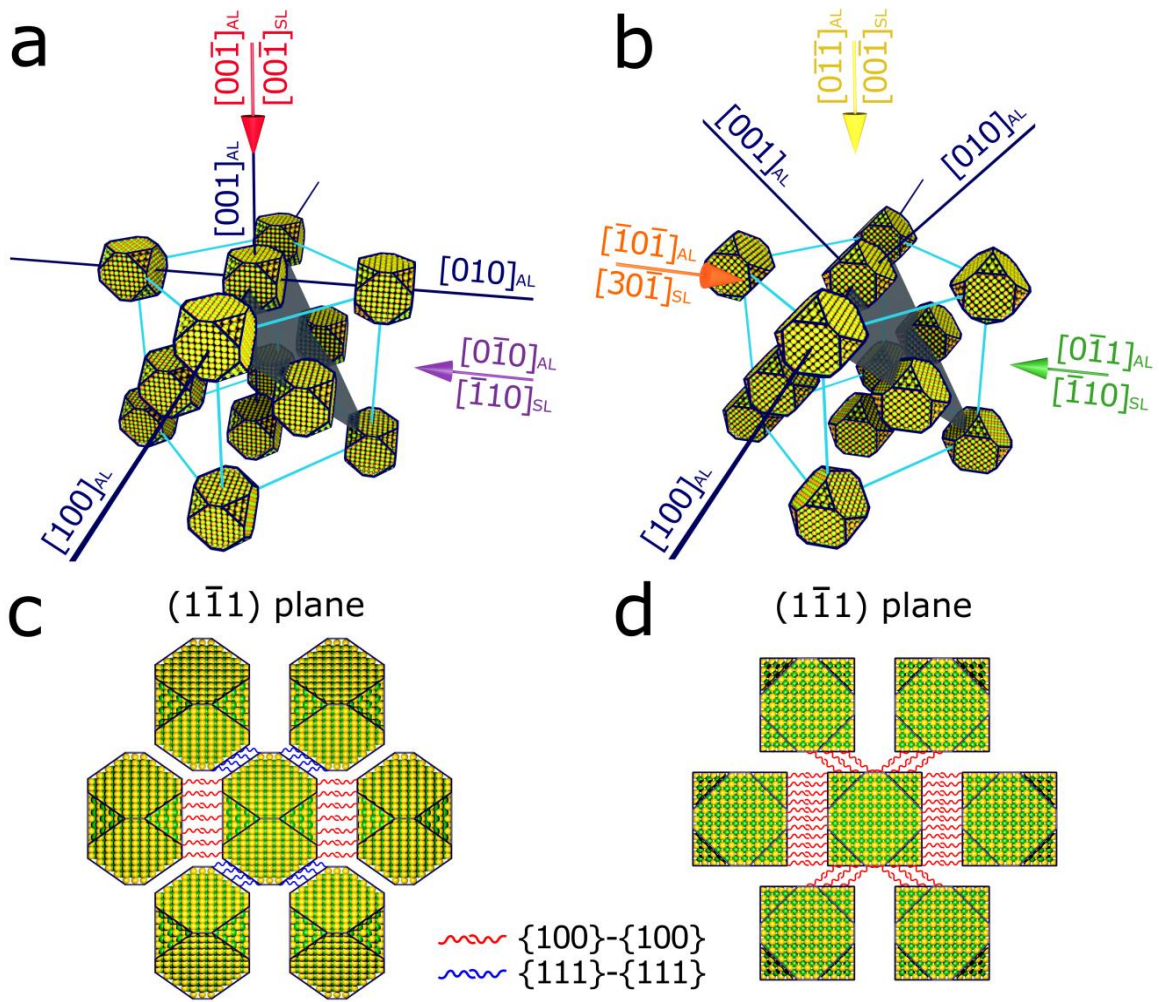

**Figure 5.** (**a,b**) Two different mesocrystalline configurations found in this work. Blue lines show the families of <100>$_{AL}$ directions of the NCs. $(1\bar{1}1)_{SL}$ planes with the highest packing density of the NCs are shown in gray. Colored arrows display the direction of the incoming x-ray beam for each domain corresponding to the diffraction patterns in **Figure 1**. (**c, d**) View along the $[0\bar{1}0]_{AL}$ (**c**) and $[0\bar{1}\bar{1}]_{AL}$ (**d**) directions of the $(1\bar{1}1)_{SL}$ planes shown in **a** and **b**, respectively. In each $(1\bar{1}1)_{SL}$ plane, six nearest neighbor NCs out of twelve are visible. The facet-facet interaction between central particle and other particles are shown by red ({100}-{100} interactions) and blue ({111}-{111} interactions) lines.

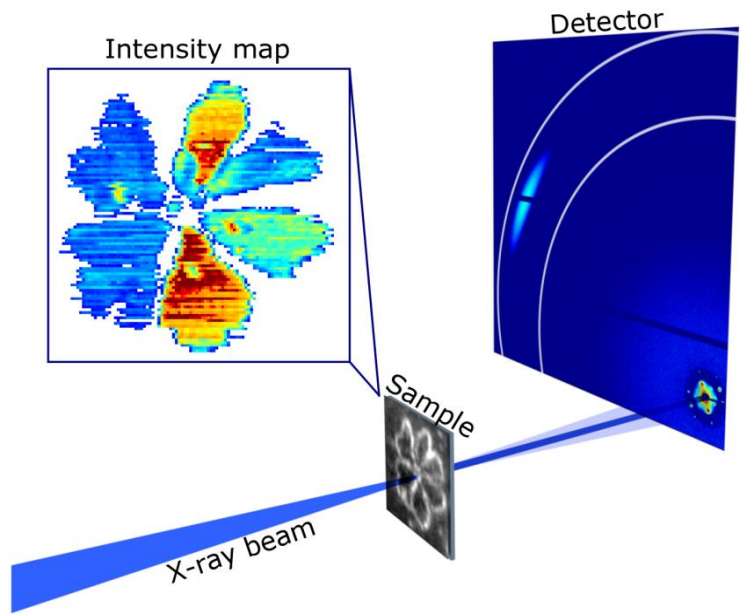

**Figure 6.** Scheme of the diffraction experiment. The sample is scanned by a nanofocused X-ray beam with the size of $400 \times 400$ nm$^2$ in transverse direction. The detector is positioned 41 cm downstream from the sample in transmission geometry and is shifted from the optical axis of the beam to measure simultaneously SL and AL Bragg peaks. The scanned map of the WAXS peak intensity between two lines marked on the detector is shown as an inset.

# Supporting Information

**Comparison of the determined lattice constant with the NC diameter**

The size of the PbS NCs was estimated by analysis of the absorption spectra, shown in **Figure S1**. According to the empirical formula, the first excitonic transition at 730 meV corresponds to the average particle diameter of 6.8 nm.[1] This value is also corroborated by the transmission electron micrograph in **Figure S2**.

The nearest-neighbor distance (NND) in an $fcc$ lattice is $NND = \frac{\sqrt{2}}{2} a$, which gives 10.6 nm for an average lattice constant a=150 Å. This value is the effective diameter consisting of one whole NC plus two layers of oleic acid. Given that the head-to-tail length of oleic acid is roughly 1.9 nm (from universal forcefield calculation), one would expect 6.8 nm + 2 · 1.9 nm = 10.6 nm for the NND from optical spectroscopy, which is in excellent agreement with the result obtained from X-ray scattering and fitting the data to an $fcc$ lattice.

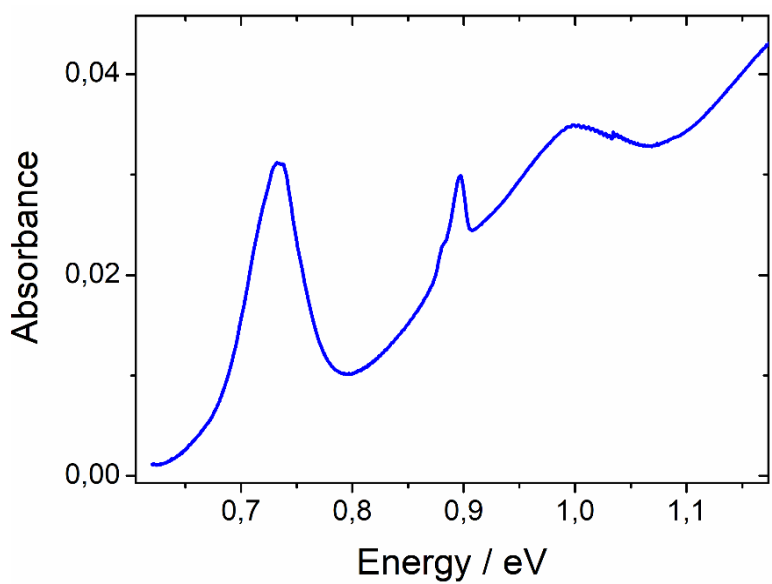

**Figure S1.** Absorption spectrum of the PbS NCs used in this study.

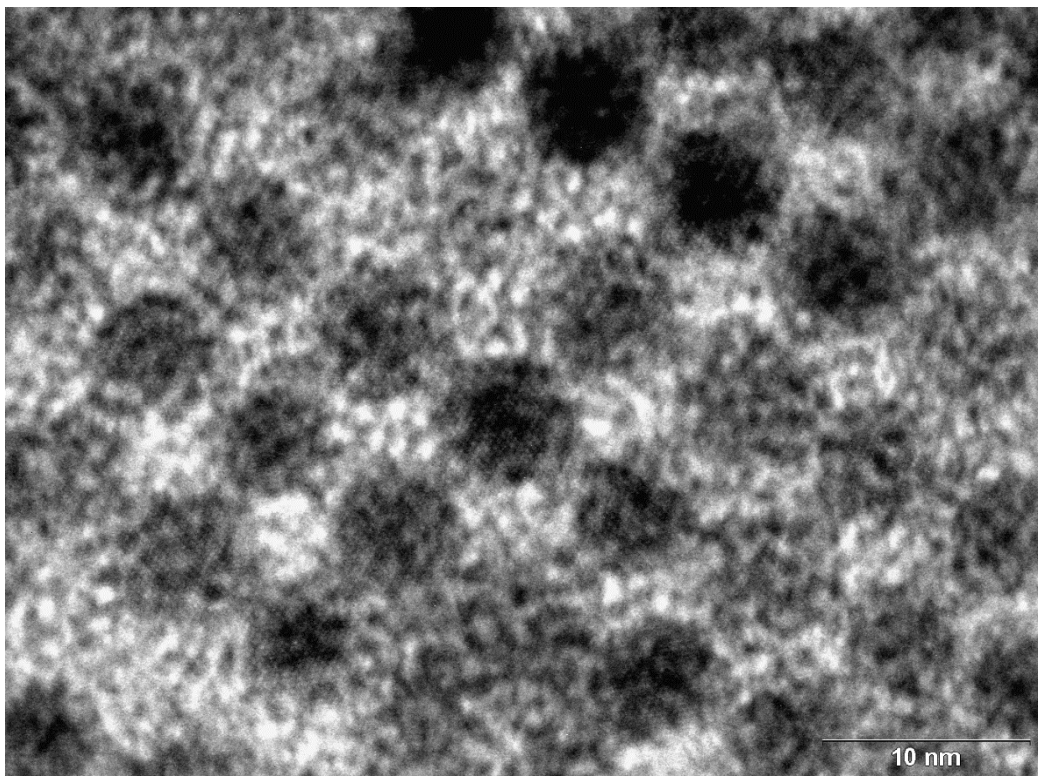

**Figure S2.** Transmission electron micrograph of the PbS NCs used in this study.

**Resolving the domain structure**

The spatially resolved maps showing the domain structure of the sample (**Figure 1a,b**) were obtained as follows. For each diffraction pattern, corresponding to a certain illuminated area of the sample, we found angular positions of the most intense peaks at $q^{SL} = 0.083$ Å$^{-1}$ and $q^{AL} = 2.12$ Å$^{-1}$ corresponding to reflections from SL and AL, respectively. The spatial points of the raster scan were classified in five different classes by the positions of the SAXS and WAXS peaks independently. Classes were designated with colors corresponding to the typical peak positions as shown in **Figure 1d-m**. Patterns which could not be attributed to any of these classes (containing peaks at positions corresponding to different classes simultaneously) were designated with gray color. Each spatial position of the maps in **Figure 1a,b** is shown with the color of the class it belongs to. White color stands for the patterns where no peaks were observed.

**Superlattice structure**

In order to determine the superlattice structure, we analyzed SAXS areas of the integrated diffraction patterns from the domains defined in **Figure 1** of the main text (shown in **Figure S3a-e**). The calculated average radial intensity of the diffraction patterns within each domain is shown in **Figure S3f-j**, where the q values are noted for the most prominent scattering peaks. The radial average of the WAXS region contains peaks at 1.84 Å$^{-1}$ and 2.12 Å$^{-1}$, which are attributed to the $111_{AL}$ and $200_{AL}$ reflections of PbS, respectively.

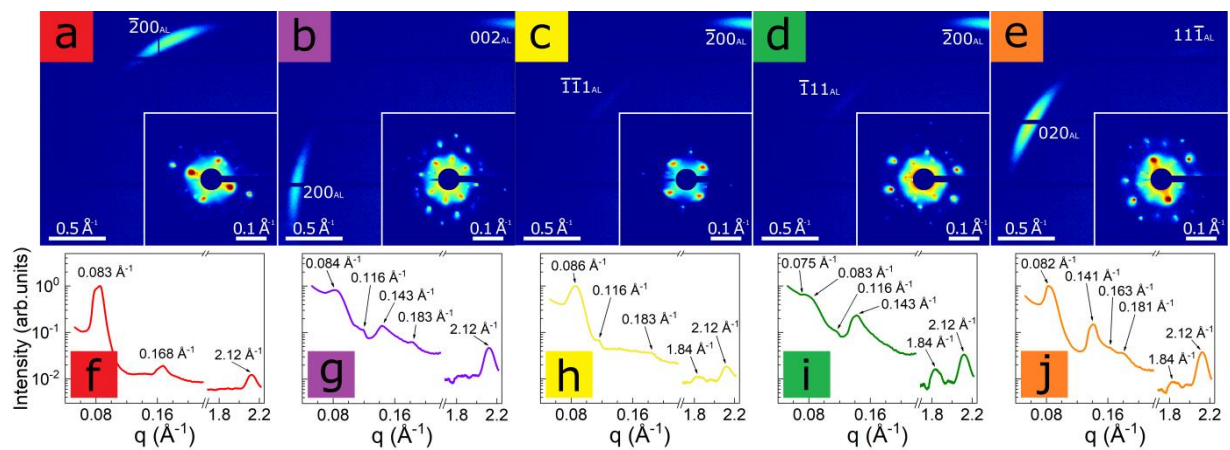

**Figure S3. (a-e)** Averaged scattering intensity of all diffraction patterns in each of the five domains defined in **Figure 1a** (color code is the same). Enlarged SAXS pattern is shown in the lower right corner (Note: The center of WAXS pattern does not coincide with the center of SAXS pattern due to a different scale). (**f-j**) Angular averaged scattering intensities of the patterns shown in the top row. The radial positions of the diffraction peaks are indicated by black arrows.

The positions of the SAXS peaks observed in the radial average for all domains are summarized in **Table S1** and indexed according to a face-centered cubic (*fcc*) lattice. The corresponding unit cell parameter *a* for each reflection is also shown in the table. An averaged unit cell parameter <*a*> slightly varies between different domains, but all values are within the standard deviation of the mean value. Thus, the superlattice structure can be described with high confidence as a face-centered cubic lattice with lattice constant $a = 150.1\pm3.0$ Å for all domains.

**Table S1**. Determination of the lattice constant in all five domains in **Figure 1** based on the *fcc* lattice.

| Domain | $q_{exp}$, Å$^{-1}$ | h | k | l | a, Å | <a>, Å |
|---|---|---|---|---|---|---|
| Red | 0.083 | 2 | 0 | 0 | 151.4 | 150.5±0.9 |
| | 0.168 | 4 | 0 | 0 | 149.6 | |
| Purple | 0.084 | 2 | 0 | 0 | 149.6 | 149.9±2.8 |
| | 0.116 | 2 | 2 | 0 | 153.2 | |
| | 0.143 | 3 | 1 | 1 | 145.7 | |
| | 0.183 | 3 | 3 | 1 | 151.3 | |
| Yellow | 0.086 | 2 | 0 | 0 | 146.1 | 150.2±3.0 |
| | 0.116 | 2 | 2 | 0 | 153.2 | |
| | 0.183 | 3 | 3 | 1 | 151.3 | |
| Green | 0.075 | 1 | 1 | 1 | 145.1 | 148.9±3.5 |
| | 0.083 | 2 | 0 | 0 | 151.4 | |
| | 0.116 | 2 | 2 | 0 | 153.2 | |
| | 0.143 | 3 | 1 | 1 | 145.7 | |
| Orange | 0.082 | 2 | 0 | 0 | 153.2 | 151.3±3.0 |
| | 0.141 | 3 | 1 | 1 | 146.5 | |
| | 0.163 | 4 | 0 | 0 | 154.2 | |
| | 0.181 | 3 | 3 | 1 | 151.3 | |

Diffraction patterns averaged over individual domains are shown in **Figure S4a-e**. All peaks are indexed assuming the incident X-ray beam to be along the $[00\bar{1}]_{SL}$ direction for the Yellow domain, along the $[\bar{1}10]_{SL}$ direction for the Green domain, along the $[30\bar{1}]_{SL}$ direction for the Orange domain, along the $[00\bar{1}]_{SL}$ domain for the Red domain, and along the $[\bar{1}10]_{SL}$ direction for the Purple domain. To confirm the suggested structure and orientations of the domains, we simulated the diffraction patterns of the superlattice as shown in **Figure S4f-j**. In these simulations, we considered the *fcc* superlattice with the experimentally obtained unit cell parameter a = 150 Å, and the form factor of the NCs in nodes of the superlattice, which were approximated by spheres with normally distributed radii (65 Å mean radius and 5 Å standard deviation).

The determined orientations of the superlattice and the NCs with respect to the incident X-ray beam allowed us to reveal the relative orientation of the NCs in the superlattice unit cell. Modelled unit cells for all five domains are shown in **Figure S4k-o**), where the direction of the incident X-ray beam is perpendicular to the plane of the figure.

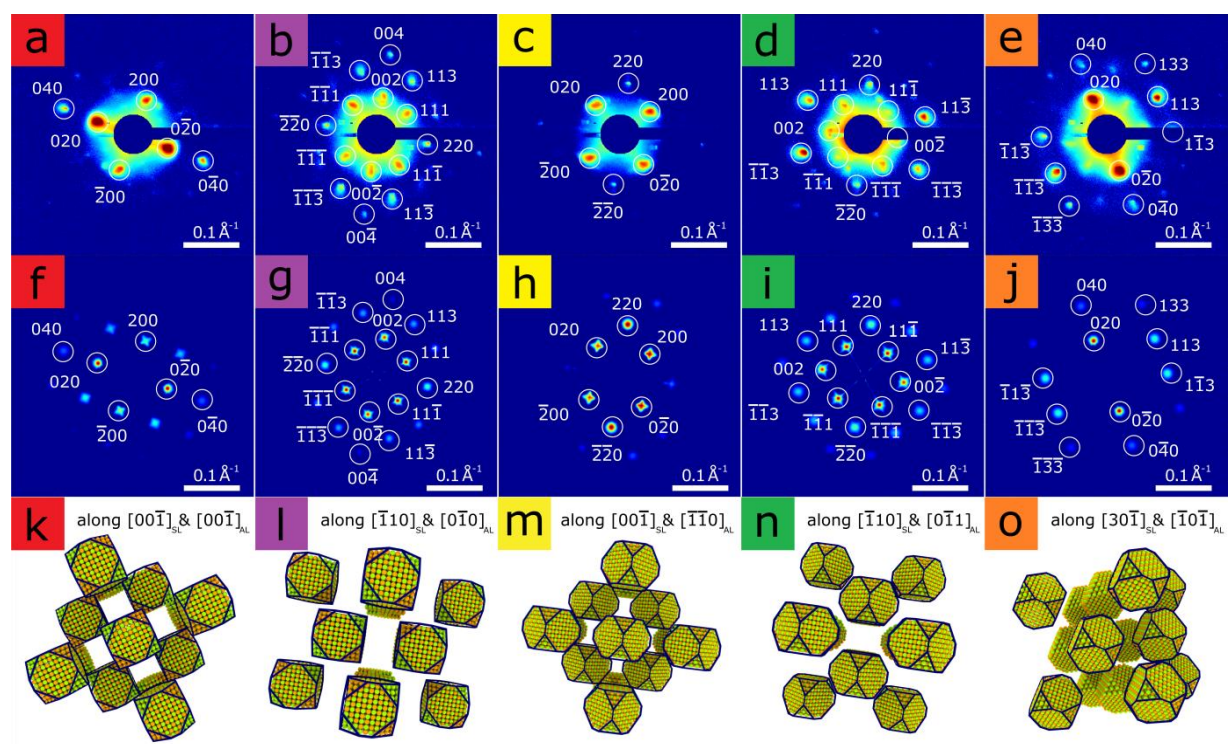

**Figure S4.** (**a-e**) Averaged scattering intensity of all diffraction patterns in each of the five domains defined in **Figure 1a** (color code is the same), all Bragg peaks are indexed according to an *fcc* structure; (**f-j**) Simulated diffraction patterns for all five domains; (**k-o**) Real-space models of the superlattice unit cell for all five domains. The view is along the incident X-ray beam.

**X-ray cross-correlation analysis**

The XCCA method is widely used for the analysis of disordered or partially ordered systems such as colloids, liquid crystals, polymers e*tc*. It is capable of providing insights on hidden symmetries, such as bond-orientational order or partial alignment of particles in the system. This method was also shown to be highly useful to study the angular correlations in mesocrystals.[2] While details and mathematical background on this method could be found elsewhere, here we briefly summarize the main concepts.[3,4]

XCCA is based on the analysis of a two-point angular cross-correlation function (CCF) that can be calculated for each diffraction pattern as

$$C(q_1, q_2, \Delta) = \frac{1}{2\pi} \int_{-\pi}^{\pi} I(q_1, \varphi) I(q_2, \varphi + \Delta) d\varphi \qquad (1)$$

where $I(q,\varphi)$ is the intensity of diffraction pattern at a certain position of the momentum transfer vector with its radial q and angular $\varphi$ coordinates. All values used in this definition are shown in **Figure S5**.

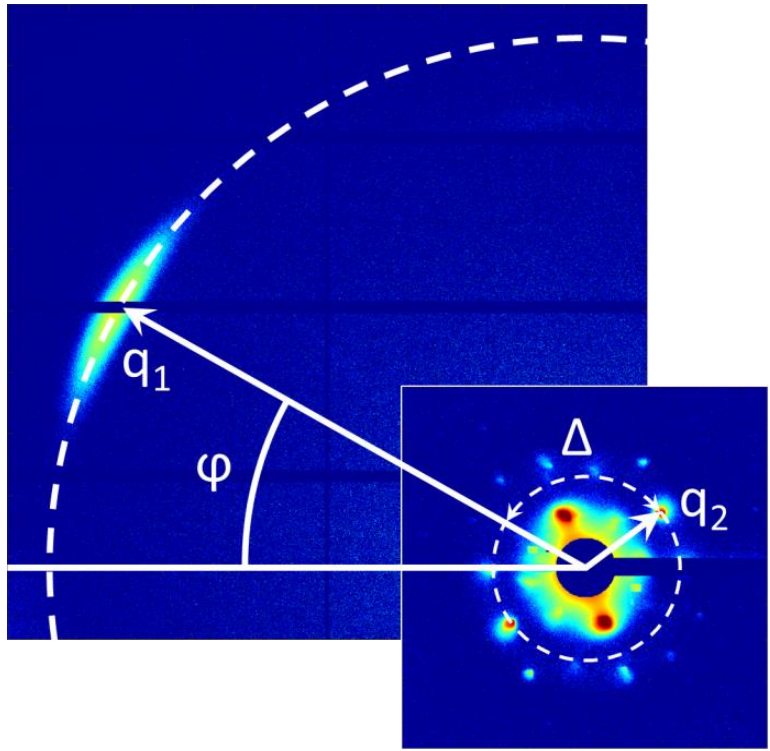

**Figure S5.** Outline of the CCF calculation. White arrows point to the Bragg reflections from the PbS AL and SL with momentum transfer values from the center of the pattern $q_1$ and $q_2$, respectively. The angle $\Delta$ between these Bragg peaks is shown. SAXS area is enlarged for better visibility.

Experimentally obtained diffraction patterns contain features that block the scattering signal, such as detector gaps, beamstop, beamstop holder, etc. In order to take their presence into account, we introduce into **Eq. (1)** the mask function

$$W(q,\varphi) = \begin{cases} 0, & \text{gaps, beamstop etc.} \\ 1, & \text{otherwise} \end{cases}. \quad (2)$$

This yields the final form of the CCF as follows:

$$C(q_1, q_2, \Delta) = \int_{-\pi}^{\pi} I(q_1,\varphi)W(q_1,\varphi)I(q_2,\varphi+\Delta)W(q_2,\varphi+\Delta)d\varphi \quad (3)$$

which afterwards can be scaled in such a way that its value lies between 0 and 1. Taking the values of $q_1$ and $q_2$ indicated in the main text, we studied the correlations between reflections in the WAXS and SAXS areas. To obtain statistically meaningful data, CCFs were averaged over all diffraction patterns from different spatial positions within each domain of the sample.

The CCF functions were simulated on the basis of the determined real-space structures for all domains. Angular positions of the SAXS/WAXS Bragg peaks $\varphi_i$ were modelled according to the determined orientations shown in **Fig. S2k-o.** The Bragg peaks in both WAXS and SAXS areas were assumed to have Gaussian shapes in the angular direction, and intensity on the corresponding ring was calculated as follows:

$$I(\varphi) = \sum_i A_i \cdot \exp\left[-\frac{(\varphi-\varphi_i)^2}{2\sigma^2}\right] \quad (4)$$

where $A_i$ is the amplitude of the *i*-th Bragg peaks in the SAXS/WAXS area from the experimental data $\sigma$ is the angular size of the *i*-th SAXS/WAXS peak. The respective angular position of the SAXS/WAXS peaks were slightly varied in order to achieve better agreement of simulated and experimental data, and the small misorientation of the SL and AL was obtained. The angular sizes of the SAXS/WAXS peak $\sigma$ were chosen to fit experimental data.

In the experiment, we were able to measure simultaneously the signal in the SAXS area for all azimuthal angles, however in the WAXS area we were limited by the detector size and measured the scattering signal only in the angular range of approximately 90° azimuthally. To simulate the effect of finite detector size we included the missing angular range for the WAXS signal

$$W(\varphi) = \begin{cases} 1, & \varphi_1 < \varphi < \varphi_2 \\ 0, & \text{otherwise + gaps, beamstop} \end{cases} \quad (5)$$

where $\varphi_1$ and $\varphi_2$ are boundary angle values of the WAXS area fitting into the detector.

The simulated CCFs were evaluated as

$$C_{sim}(q_1, q_2, \Delta) = \int_{-\pi}^{\pi} I_{SAXS}(q_1, \varphi) I_{WAXS}(q_2, \varphi + \Delta) W(\varphi + \Delta) \, d\varphi \qquad (6)$$

to take effects of the mask into account, and then normalized to values between 0 and 1 as described above.

**NCs orientational disorder**

XCCA analysis shows that in average the individual nanocrystals are aligned in a certain way with respect to the crystallographic directions of the superlattice. The small deviations in orientation of NCs from this alignment can be described in terms of the misorientation angle $\Delta\Phi$. In order to obtain a statistical description of $\Delta\Phi$, we considered the average diffraction pattern of each WAXS domain. The enlarged part of the diffraction pattern averaged over the red domain, which contains $200_{AL}$ peak, is shown in **Figure S6a**. The radial and azimuthal cross sections of the averaged WAXS peak are shown in **Figure S6b,c**. We find that both cross sections are well-fit by Gaussian function, wherein the width of the diffraction peak in azimuthal direction is significantly larger than in the radial direction.

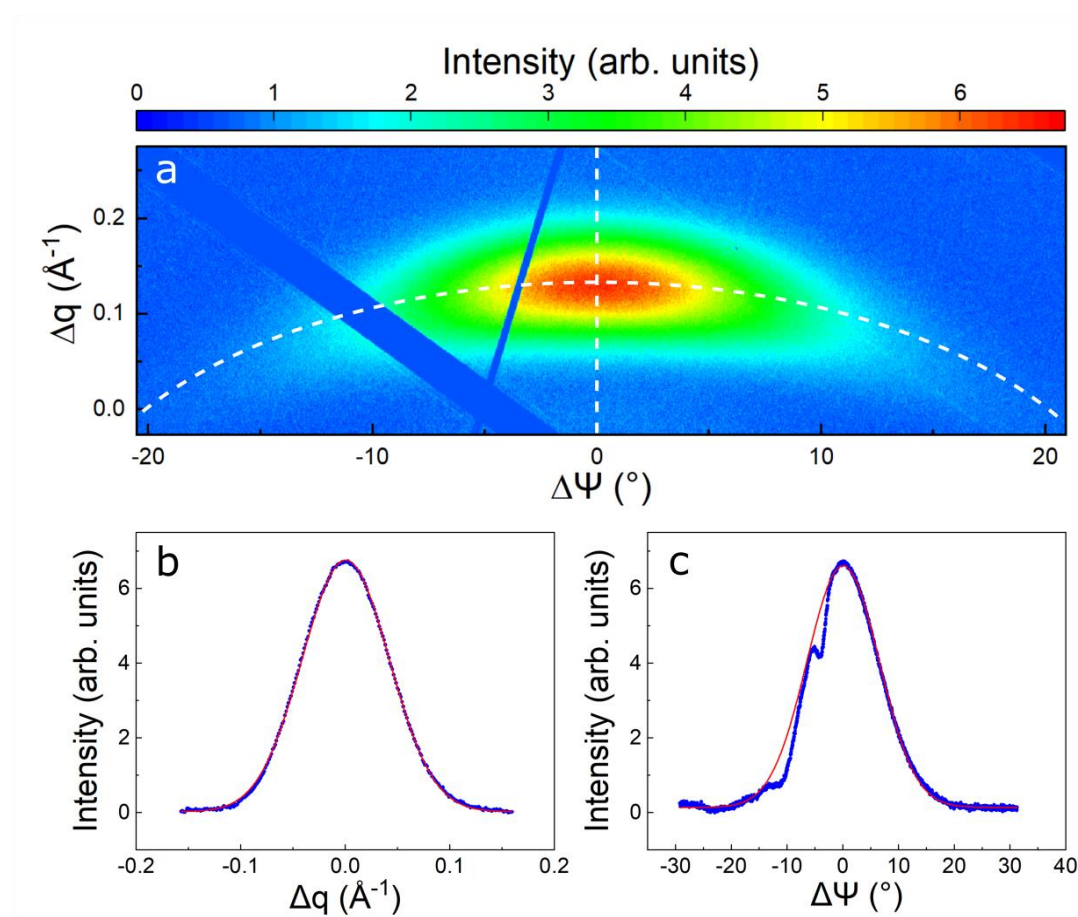

**Figure S6. (a)** Two-dimensional shape of the $200_{AL}$ diffraction peak averaged over diffraction patterns from the red domain. White dashed lines represent cross sections in radial and azimuthal directions. Radial (**b**) and azimuthal (**c**) profiles of the 200 AL peak, respectively. Blue points are experimental data, and red lines are Gaussian fits.

To quantitatively characterize angular disorder we assumed that WAXS peak broadening is caused by two factors: Sherrer broadening due to the small size of the NCs and orientational disorder of the NCs in sites of the SL. The first factor affects both radial and azimuthal width of the peak while the second one influences only the azimuthal width of the peak. Assuming that these two factors are independent, we can estimate the value of the orientational disorder $\Delta\Phi$ from the relationship between radial and azimuthal widths of the peak:

$$\sigma_{az}^2 = \sigma_{rad}^2 + \Delta\Phi^2 \qquad (7)$$

where $\sigma_{az}$ and $\sigma_{rad}$ are FWHM values of the peak angular size in azimuthal and radial directions obtained from the gaussian fitting, respectively.

Both angular and radial FWHM values calculated for all the domains are summarized in **Table S2** together with evaluated values of the orientational disorder. Thus, the orientational disorder ($\Delta\Phi$) of the almost iso-oriented atomic lattices of NCs is roughly 15° in all domains, similar to the recently reported $\Delta\Phi$ for superlattices of tetrathiafulvalene-linked PbS NCs.[2]

**Table S2.** Determination of the orientational disorder in five domains.

| Domain | Azimuthal FWHM,° | Radial FWHM,° | $\Delta\Phi$,° |
|---|---|---|---|
| Red | 13.1±0.5 | 2.5±0.1 | 12.9±0.6 |
| Purple | 17.5±1.2 | 2.5±0.1 | 17.3±1.2 |
| Yellow | 16.9±0.8 | 2.5±0.1 | 16.7±0.8 |
| Green | 16.7±1.5 | 2.6±0.1 | 16.5±1.4 |
| Orange | 13.1±0.5 | 2.4±0.1 | 12.9±0.6 |

**Compression of the superlattice at the grain boundaries**

For each spatial position of our sample, we estimated the radial positions of $<311>_{SL}$ family of peaks. The calculated radial average for each diffraction pattern was fitted with a Gaussian function in the range of magnitudes of the scattering vector from 0.130 to 0.160 Å$^{-1}$. Resulting positions of the peaks are shown in **Figure S7a** in the false color code. One can see the magnitude of the scattering vector corresponding to the $<311>_{SL}$ peaks tends to increase while approaching the borders of the domains, similarly to the $<200>_{SL}$ family of peaks (see **Figure 3** in the main text).

To verify our findings, we fitted radial averages more precisely with appropriate background subtraction for the grain boundary shown as scan S10 in **Figure S7a.** The values of the momentum transfer corresponding to the $<311>_{SL}$ peaks are shown in **Figure S7b**. Distances are relative to the edge of the domain (defined by the absence of the SAXS diffraction). The length of the scattering vector increases from about 0.141 Å$^{-1}$ in the bulk of the domain up to about 0.162 Å$^{-1}$ on the very edge that gives 13% shrinkage of the unit cell (see also **Figure 3** of the main text).

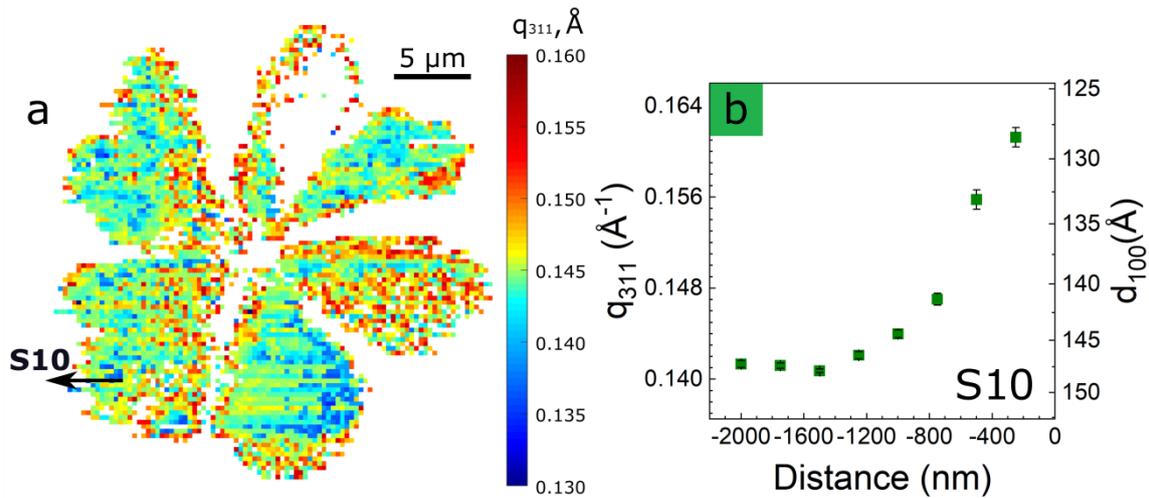

**Figure S7**. (**a**) Spatially resolved map of the scattering vector length for peaks the Bragg peaks of the $<311>_{SL}$ family. The color code quantifies variations in q. The black arrow refers to specific scan, which is shown on the right panel. (**b**) Spatial variation in $q_{311}^{SL}$ and calculated $d_{100}^{SL}$-spacing during scanning from the bulk towards the edge of domains shown as S10.

**Rotation of the superlattice near domain boundaries**

We analyzed Scans S1 and S4-S11 shown in **Figure 1** in the same way as in the main text. We found that the superlattice rotates by different angles at the grain boundaries. It is important to note that the length of the transition region is different in each case. We present results of our analysis in two groups: one in **Figure S8** where diffraction patterns are taken with the step size of 500 nm and another group in **Figure S9** where diffraction patterns are taken with the step size of 1000 nm.

In all analyzed scans, the relative intensity of the SAXS peaks changes and some new peaks emerge. In all cases the evolution of the SAXS pattern along the scan can be explained with a continuous rotation of the unit cell in the range of 8-14°. Approximate directions and angles of rotations within the scans are indicated on the left panel of Figure **S8** and **S9**.

A similar analysis of the AL rotation is complicated due to the fact that the rotation angle of the AL unit cell lies within the range 8-14° for scans S2, S4-S11 (see **Table S3**), and these changes are difficult to track on top of the orientational disorder of the NCs, which is roughly 15°. For this reason, scan S1 as the most obvious example for the rotation of the atomic lattice near the grain boundaries is shown in the main text. However, in all other scans the gradual extinction of the WAXS peaks supports our hypothesis that NCs are simultaneously rotating with the superlattice (clearly seen in **Figure S8a-c**, for example).

**Table S3**. Rotation angles for 9 scans through the domain boundaries.

| Scan  | S1  | S4 | S5  | S6  | S7  | S8  | S9 | S10 | S11 |
|-------|-----|----|-----|-----|-----|-----|----|-----|-----|
| Angle | 14° | 8° | 10° | 14° | 10° | 12° | 9° | 11° | 9°  |

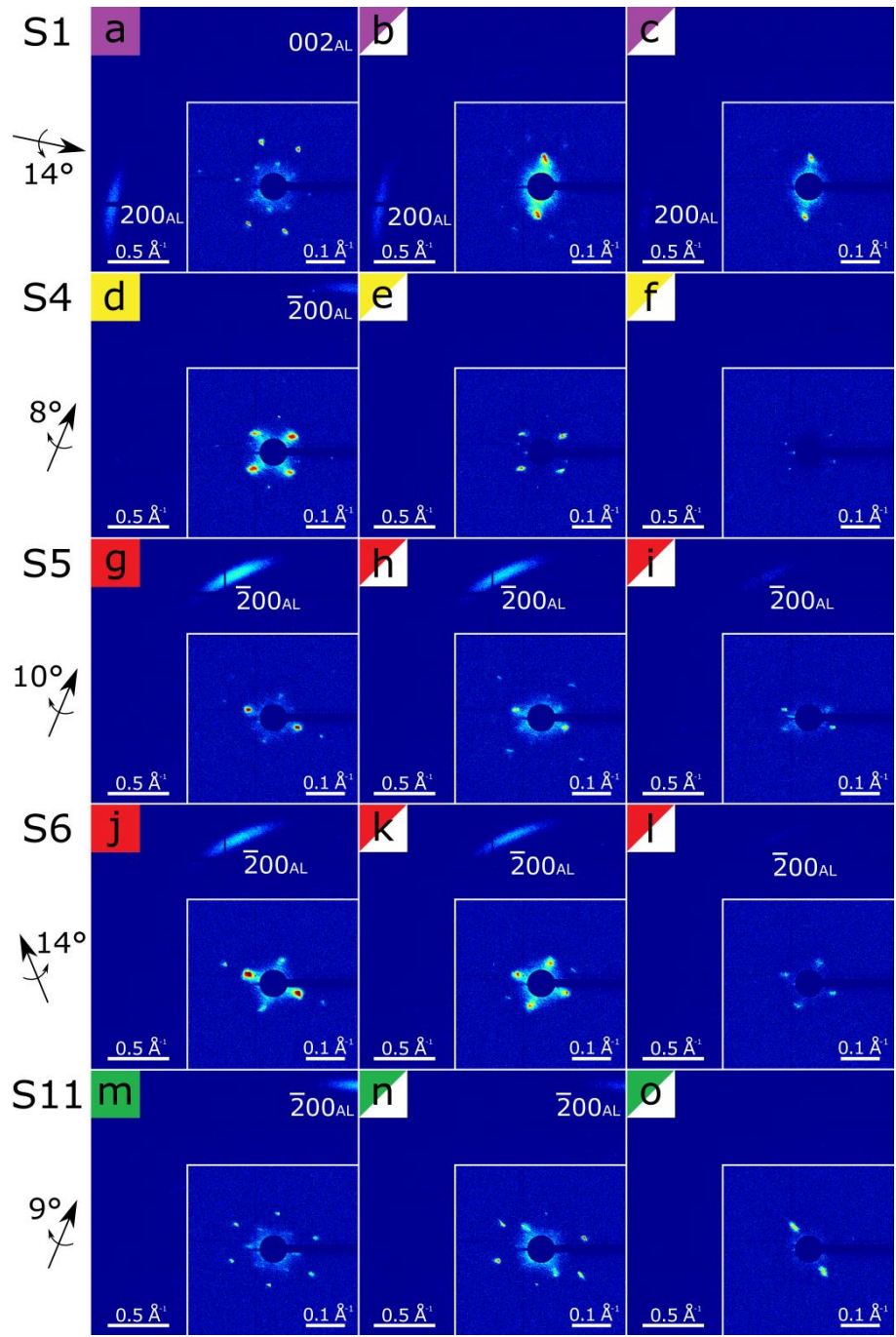

**Figure S8**. Scans through the domain boundaries with 500 nm step size. The respective scan numbers are shown on the left.

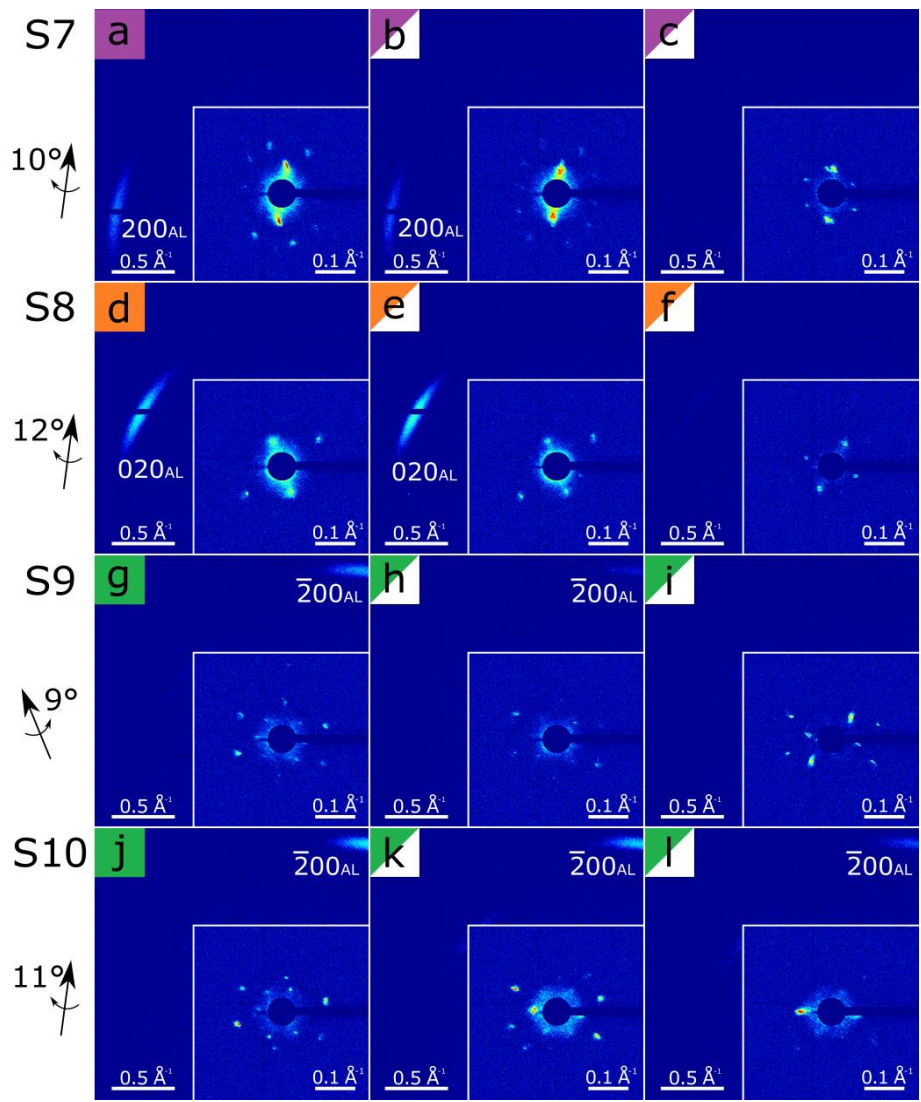

**Figure S9.** Scans through the domain boundaries with 1000 nm step size. The respective scan numbers are shown on the left.

Results of simulation of the lattice rotation in Scan S2 are presented in **Figure S10**. Simulated diffraction patterns correspond to the SL oriented along $[30\bar{1}]_{SL}$ (**Figure S10d**), $[12\ 1\ \bar{7}]_{SL}$ (**e**) and $[10\ 1\ \bar{9}]_{SL}$ (**f**) axes.

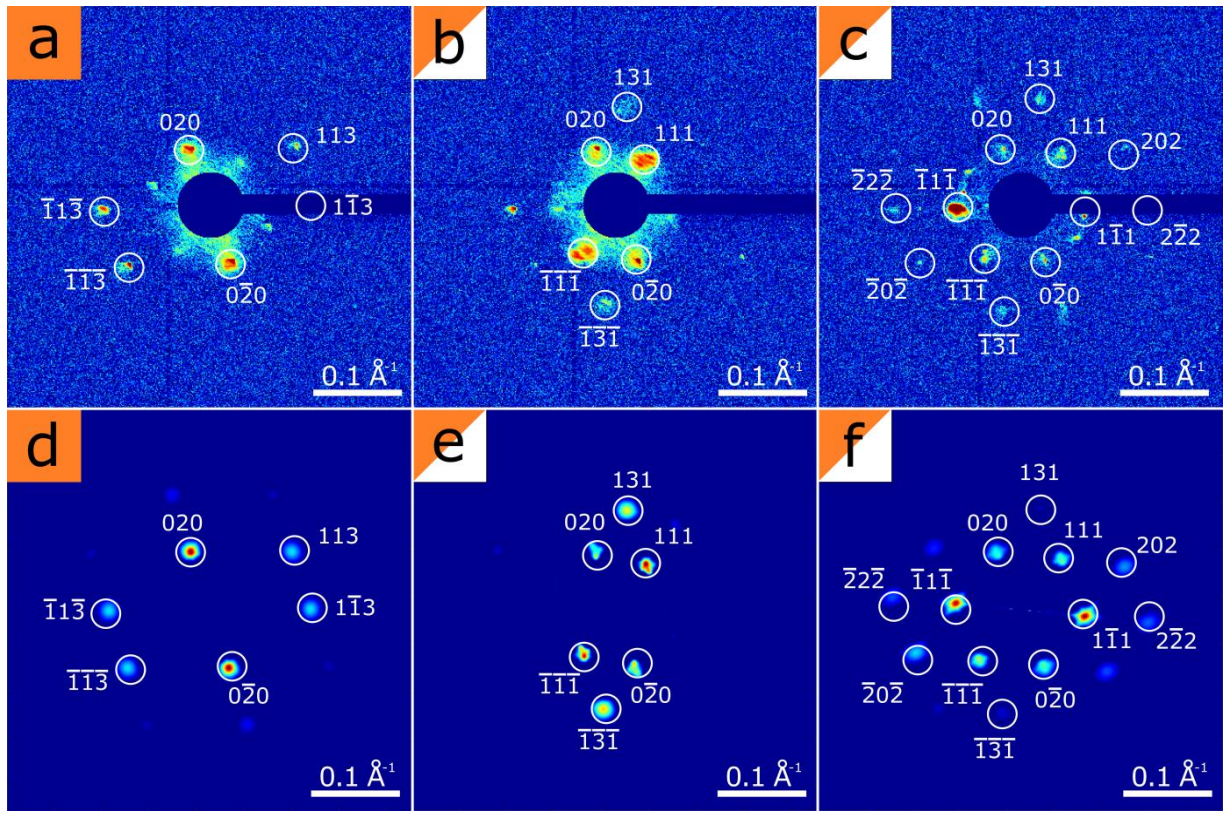

**Figure S10.** (**a-c**) Experimental and (**d-f**) simulated SAXS diffraction patterns from the Scan S2 from **Figure 1a,b** of the main text. The simulation starts from the SL oriented along $[30\bar{1}]_{SL}$ (**d**), and close to the edge of the domain superlattice is continuously rotated around $[010]_{SL}$ axis.

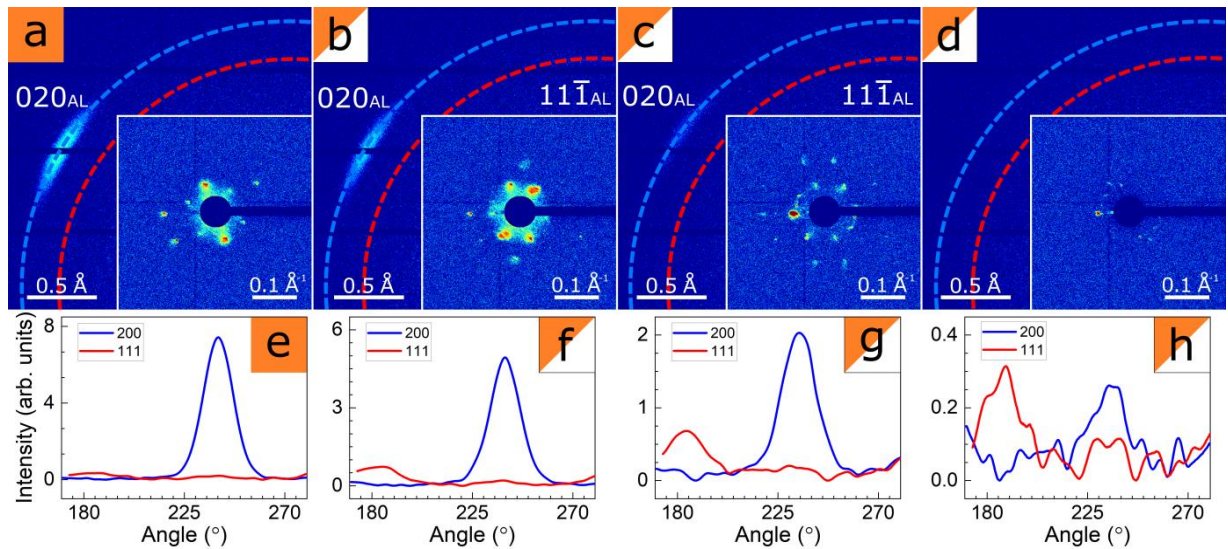

**Figure S11.** (**a-d**) Experimental diffraction patterns from the Scan S2 from **Figure 1a,b** of the main text. The analyzed q-values are indicated by the blue dashed line for $q_{020}^{AL}$ = 2.12 Å$^{-1}$ and red dashed line for $q_{111}^{AL}$ = 1.84 Å$^{-1}$. **e-h)** Azimuthal profiles of the *q*-rings corresponding to $q_{020}^{AL}$ = 2.12 Å$^{-1}$ (blue) and $q_{111}^{AL}$ = 1.84 Å$^{-1}$ (red). Changing the relative intensity of [020]$_{AL}$ and [11$\bar{1}$]$_{AL}$ indicates the rotation of the AL close to the edge of the domain.

**Facet-induced interactions stabilizing structural configurations**

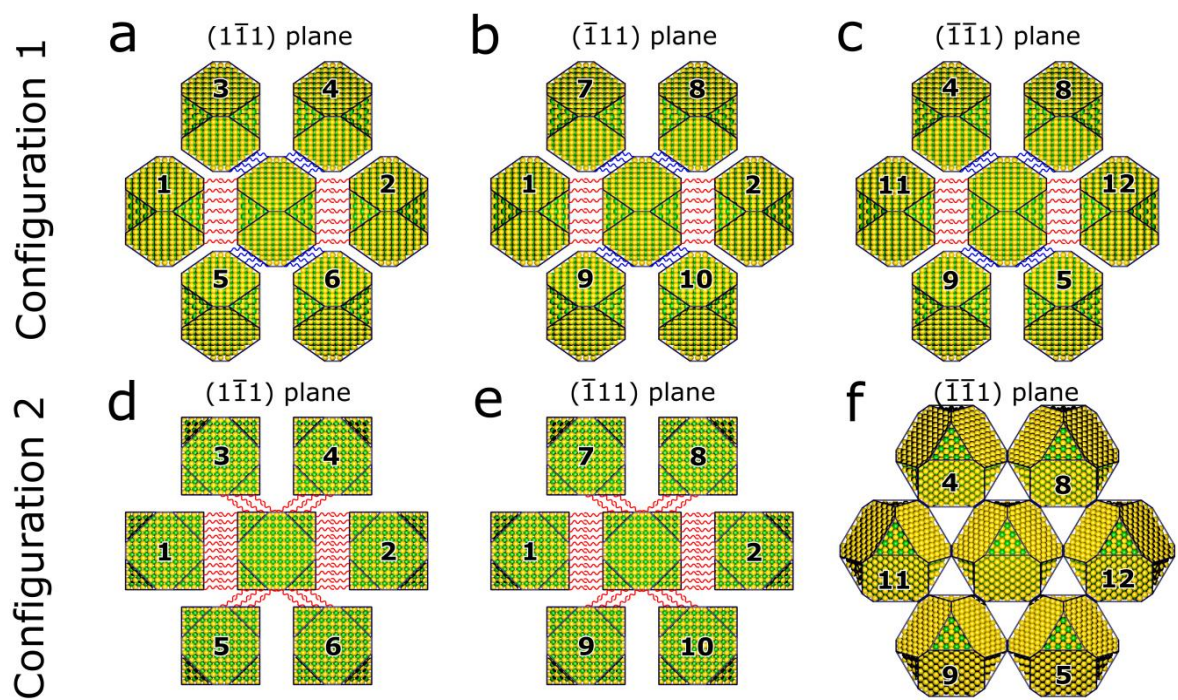

**Figure S12.** Three different {111} planes for each configuration shown in **Figure 5** of the main text. In each plane 6 nearest neighbor NCs out of 12 are visible. All 12 of the nearest neighbors of the central nanocrystal are numbered. The facet-facet interaction between central particle and other particles are shown by red ({100}-{100} interactions) and blue ({111}-{111} interactions) lines.